  \documentclass[12pt,preprint]{aastex}
 \usepackage{ifpdf}
 \usepackage{epsfig}
  \usepackage[latin1]{inputenc}
 \usepackage{amsfonts}
 \usepackage{amsthm}
 \usepackage{amssymb, latexsym, amsmath}
 \usepackage{amsbsy}
\usepackage{amstext}
\usepackage{amsopn}
\usepackage{amsgen}
\usepackage[dvips]{color}
 \usepackage{graphicx}
  \usepackage{color}
\newcommand{\eb}{\begin{equation}}
\newcommand{\ee}{\end{equation}}
\newcommand{\st}{KIC 5773205}
\shorttitle{Eclipsing M dwarf with commensurability}
\shortauthors{Makarov \& Goldin}
\begin{document}
\title{A $5/4$ commensurability of KIC 5773205, the smallest eclipsing red dwarf detected by the Kepler mission}
\author{Valeri V. Makarov}
\email{valeri.makarov@navy.mil} 
\affil{US Naval Observatory, 3450 Massachusetts Ave NW, Washington DC 20392-5420, USA}
\author{Alexey Goldin}
\email{alexey.goldin@gmail.com} 
\affil{Teza Technology, 150 N Michigan Ave, Chicago IL 60601, USA}

\date{Accepted . Received ; in original form }

\label{firstpage}
\begin{abstract}
KIC 5773205 is the least luminous eclipsing M dwarf found in the Villanova catalog of eclipsing binaries detected by the
{\it Kepler} mission. We processed and analyzed the three available quarters of mission data for this star and discovered
a persistent periodic variation of the light curve with a period, which is in exact 4:5 commensurability to the orbital
period. Three routes of interpretation are considered: 1) non-radial pulsations excited by the tidal interaction
at a specific eigenfrequency; 2) a high-order spin-orbit resonance caused by the tides; 3) an ellipsoidal deformation
caused by an outer orbiting companion in a mean motion resonance. All three explanations meet considerable
difficulties, but the available facts seem to favor the tidally driven pulsation scenario. The star may represent a new type of
heartbeat binary with tidally excited pulsations that are close to the orbital motion in frequency. \end{abstract}


\section{Introduction}
The third version of the eclipsing star catalog detected by the Kepler mission includes 2787 systems
\citep{kir}. The primaries in these systems represent various luminosity classes and evolutionary states
of Galactic stellar population, including the main sequence, giants and subgiants, and hot subdwarfs (sdO
and sdB). However, the small stars that are found at the bottom of the main sequence and degenerate stars
are rare in this catalog. There are only two binaries among those present in the Gaia DR2 catalog with
absolute $G$-band magnitudes ($M_G$) fainter than 10. One of them, KIC~10544976, is a known WD$+$M binary
\citep{rit,par} with an orbital period of $0.3504691$ d \citep{sla} and interesting out-of-eclipse ellipsoidal
variations \citep{lur}. The other, \st, which is the subjects of this paper, is less well known. Its eclipse
period in \citet{kir} is 0.2668707(17) d
\footnote{Throughout this paper, numbers in round brackets stand for standard errors of values in the last
significant digits, i.e., $1.549(65)$ is equivalent to $1.549\pm 0.065$}, 
which is closer to the short end of the distribution. No secondary
eclipses are present, hinting at a considerable eccentricity. The eclipse time variation data from \citet{con} do not
show any discernible signal. 

Although M dwarfs are very common in the Galaxy and in the Solar neighborhood in particular \citep{hen}, there are not many 
known eclipsing binaries with late M dwarfs as primaries. This is explained by their relatively small size, faintness,
and the limited range of the mass function for the companion. \st\ matches the Gaia DR2 source 2103827298302304128 with
the full complement of astrometric and photometric measures. Its observed parallax $\varpi=2.83(21)$ mas,
broadband $G$ magnitude $18.7750(26)$, and the color $BP-RP=3.19(17)$ mag, identifies this star as an M4.5V
dwarf with an effective temperature of $\sim 3100$ K according to E. Mamajek's stellar data, but the estimated 
absolute luminosity $M_G\simeq 11.04$ mag suggests
a slightly larger dwarf of the M3.5V type.
Thus, the absolute magnitude seems slightly too luminous for the observed
color. This excess brightness may be attributed to the presence of another dwarf in the binary system. Generally,
the star appears at first sight to be a nondescript field dwarf at 352 pc from the Sun, lacking distinguishing features
such as enhanced X-ray or ultraviolet radiation.

The long-cadence light curve is available for three quarters of the {\it Kepler} mission: Q14, Q15, Q16. We use the raw 
Simple Aperture Photometry (SAP) Kepler light curves rather than the Pre-search Data Conditioning (PDC)
fluxes derived by the data processing pipeline \citep{chr}. To remove the slow variations of flux in the raw data,
we employ our own Principal Component
Analysis (PCA) pre-whitening and filtering procedure described in more detail in \citet{mak16, mak17}.
We always use 4 principal components for both astrometric and
photometric data divided into quarter-year segments ($\sim90$ days). The resulting cleaned light curves and astrometric
trajectories are further analyzed in this paper.

\section{Variability-induced motion (VIM) test}

Three kinds of variability are clearly present in the light curves of \st: 1) periodic eclipses caused by the binary
companion; 2) small and moderate flares; 3) a quasi-periodic signal. All these variations are also visible in the
measured photocenter coordinates. The synchronous and correlated behavior of the measurements is a natural
consequence of the first-moment estimation technique coupled with a limited sample area. Photometry and astrometry
with Kepler are interdependent because a small motion of the digital aperture (due to the differential aberration or
a pointing jitter) changes the amount of collected light, while any intrinsic variation of the stellar flux shifts the
measured photocenter. The latter is exacerbated by the low angular resolution and rather large digital apertures, which
include parts of chance neighbors' images. The fraction of blended light also depends on the orientation of the aperture
on the sky for the same configuration of stars. Therefore, the observed photocenter location changes between the quarters
by great amounts compared to the expected photon noise level. The correlated motion of photocenter centroids and variation
of flux provides an easy way of testing the origin of this variability in the simplest and most common case when one
of the neighbors dominates the blended flux. If the intended target (close to the center of the aperture) is variable and
becomes brighter, the photocenter moves toward it along the line connecting it with the dominant neighbor. If it is the
neighbor that becomes brighter, the photocenter moves in the opposite direction. Generally, the centroid moves toward
the brightening source, but the picture becomes complicated when a few competing blends are present or only some part
of the blended image is covered by the aperture.

Our comprehensive analysis of VIMs in the main mission data \citep{mak} detected two instances for \st\ in quarters Q14 and Q15
with position angles (counted counterclockwise as seen on the sky, north through east) of $280\degr$ and $267\degr$. 
This analysis for \st\ is further complicated by the fact that the ``optimal" digital aperture, where the flux was detrmined,
includes only a single pixel, which was apparently meant to lower the impact of brighter neighbors.
We implemented a much improved reprocessing of VIM effects in the long-cadence data consistently using our PCA cleaning
method and discarding all saturated objects, to be published elsewhere. This new reprocessing resulted in a confident detection
for all three available quarters: $280\degr$ in Q14, $267\degr$ in Q15, and $271\degr$ in Q16. We note a good consistency with
the previous result and persistent, even though slightly scattered, direction of the correlated astrometric wobble.
The Gaia DR2 catalog \citep{bra}, as well as the Pan-STARRS DR2 catalog and online stacked charts \citep{cha} reveal the presence
of two slightly brighter optical (unrelated) companions at separations $8\farcs893$, $12\farcs919$ and position angles
$83.5\degr$, $156.0\degr$ from our object of interest. A few much fainter objects are also present within the Kepler digital
aperture, but they hardly matter. It appears that the closer optical companion, which is brighter than the target by
$0.619$ mag in $G$, is the main source of the observed VIM, with a smaller contribution from the more distant blue
neighbor. The important fact is that these perturbing neighbors are situated in the opposite direction to the observed VIM.
This confirms that the photometric variability detected by {\it Kepler} belongs to the intended target.

\section{Photometric variability}

\begin{figure}[htbp]
\epsscale{0.5}
  \centering
  \includegraphics[angle=0,width=0.48\textwidth]{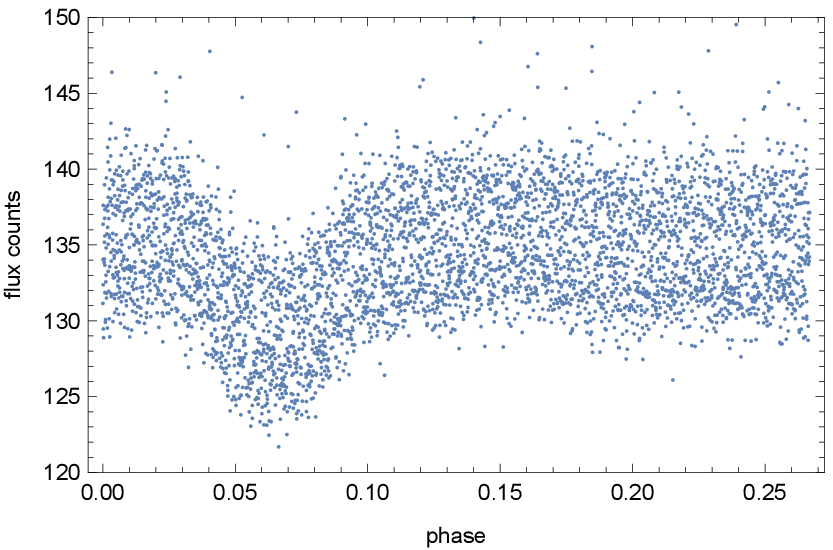}
  \includegraphics[angle=0,width=0.48\textwidth]{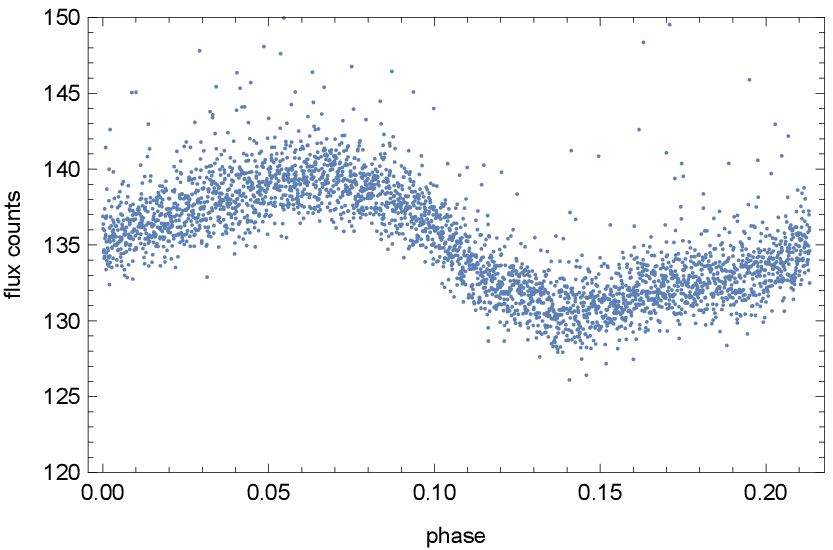}
\hspace{2pc}
\caption{Folded light curves of \st\ for quarter Q14.
Left: The complete cleaned data folded with the eclipse period $0.26687$ d and an arbitrary time zero-point.
Right: The out-of-eclipse part of the observed light curve folded with a period of $0.213236$ d.}
\label{fold.fig}
\end{figure}

The Villanova Catalog of Eclipsing Binaries lists \st\ with a period $P_o=0.2668707(17)$ d and an eclipse depth of 0.0478 \citep{kir}. 
The eclipse is clearly visible if we fold the light curve with this period, see Fig. \ref{fold.fig}, left. There is a scatter
of data points well above the folded light curve, which is caused by short-term flares of moderate magnitude. There is no obvious
periodicity or order in the occurrence of flares. There is a previously unnoticed third type of variability, which we investigate
in this paper. It causes the apparently uniform dispersion of flux measurements within $\sim 10$ electrons per second in the folded
light curve.

\begin{figure}[htbp]
\epsscale{0.75}
  \centering
  \includegraphics[angle=0,width=0.48\textwidth]{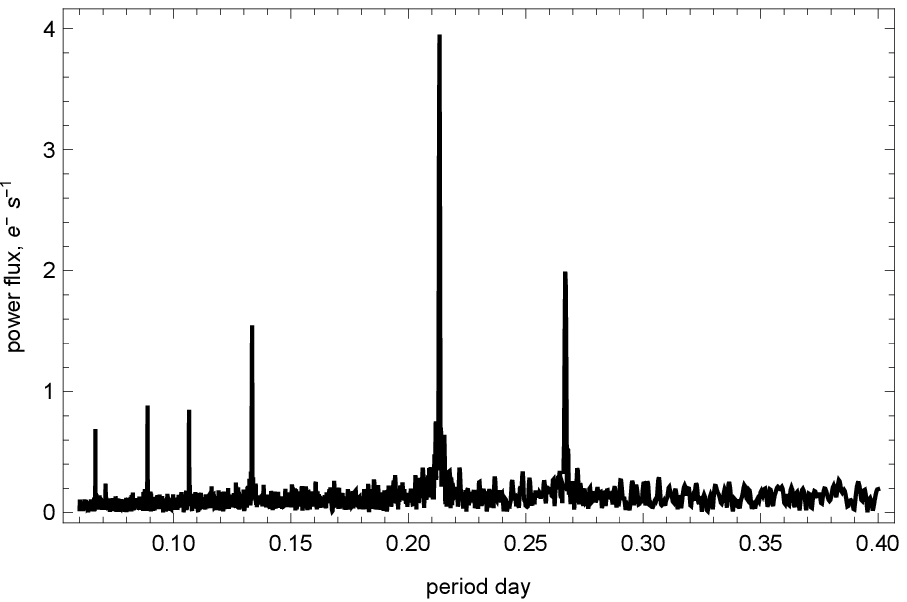}
  \includegraphics[angle=0,width=0.48\textwidth]{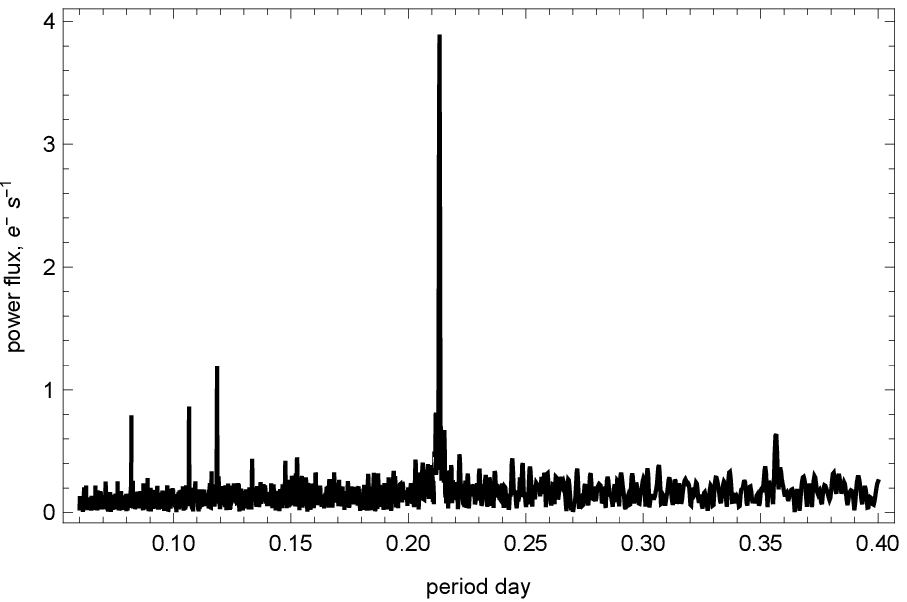}
\hspace{2pc}
\caption{Amplitude periodograms of the long-cadence light curve of \st\ in Q14. Left: The entire light curve data
are included, showing the two main periodicities and their harmonics. Right: Only the out-of-eclipse segments
are included, removing the eclipse signal but introducing aliases.}
\label{perio.fig}
\end{figure}

Fig. \ref{perio.fig} shows the standard periodogram of all out-of-eclipse flux measurements. In the left plot,
the entire collection of Q14 flux measurements is used. The peak at $P_o$ and its harmonics represent the eclipses,
but they are not the largest signals in the data. In order to remove the eclipses,
we discarded all data points within a window of $0.0678$ d centered on each eclipse minimum. This simple technique allows us to
suppress a series of signals in the periodogram caused by the eclipse of limited duration and strongly non-sinusoidal shape.
The remaining peaks in Fig. \ref{perio.fig}, right, 
reveal the presence of other periodic signals. The height of each peak is exactly the amplitude of the best-fitting
sinusoid (with a free phase) at the corresponding frequency. The most powerful out-of-eclipse sinusoidal signal has a period of
$0.2132(3)$ (in Q14) and an amplitude of $3.94$ e$^-$ s$^{-1}$. It is in fact more powerful than the eclipses. We will use $P_e$ 
to denote
this period. There are two less prominent peaks at $0.10663(7)$, $0.07107(4)$ d, with amplitudes $0.94$, $0.42$ e$^-$ s$^{-1}$, which are obviously
the second and third harmonics of the main mode, i.e., $P_e/2$ and $P_e/3$. Their presence signifies a non-sinusoidal shape of
the mode, which is also visible in Fig. \ref{fold.fig}, right. Finally, there seems to be a few other prominent signal with a periods
$P_a=0.1185(1)$ (amplitude $1.26$ e s$^{-1}$), 0.08207 d, and 0.06277 d, which are not present in the full-data
periodogram on the left. We posit that it is an artifact, which is sometimes called alias in the literature.
It is caused by the removal of segments of data with a period of $P_e$ and the interference of these blank spaces with the
actual periodicity. Indeed, we observe that $1/P_a=1/P_e+1/P_o$ to within 0.15\% of its value, and the other periods
are exactly $1/(1/P_e+k/P_o)$, $k=2,3$.

The presence of a periodic signal with $P_e$ is quite striking, because the ratio of the two physical periods, $P_o/P_e$ is equal
to 5/4 within 0.16\% of its measured value. This can hardly be attributed to chance, hence, the signals should be interrelated.
The star and the eclipsing companion are separated by roughly $1.5\,R_\sun$ (assuming a total mass of $0.4\,M_\sun$), which may be
close enough to generate tidal deformations of the star. The ellipsoidal deformation and the related gravity-dependent surface
brightness modulation emerge in light curves as a mainly semidiurnal variation, i.e., with a period of $P_o/2$. Contrary to this
expectation, we do not find any signal of note at $P_o/2$, but instead find a strong quasi-sinusoidal signal at $4\,P_o/5$. 

\section{The origin of commensurability}

As in many other cases of unclear interpretation of periodic signals observed from stars, whether photometric or spectroscopic,
there are three general routes that may be interrelated, viz., non-radial pulsation, rotation, and orbital perturbation \citep{wil,hat}.
Manifestations of these mechanisms are so similar in the limited data that any interpretation bears a good deal of uncertainty.
These mechanisms may also work together, as in the case of a spotted star synchronously rotating with a close companion. We
explore in this study the possibility that the phases and amplitudes of the observed Fourier modes can be decisive in
selecting the most likely interpretation.

To estimate the relative phase of the commensurate signals, we combine the three quarters of flux measurements into one data set
by converting each quarter's flux into relative flux deviation $\delta F=(F-F_{\rm median})/F_{\rm median}$ expressed in parts
per thousand (ppt). This step is required because of a different photometric setup between the quarters resulting in a much
different flux level. The joint three quarters of data are cleaned of the eclipse segments and then fitted with a series of
Fourier functions, which are the unity and $\sin$- and $\cos$-harmonics of $P_e$, i.e., $2\pi(t-t_0)/(k\,P_e)$, with $k=1, 2, 3$.
The 7 unknown coefficients are determined in a non-weighted least-squares solution. The amplitudes of harmonics are computed as
$\sqrt{s_k^2+c_k^2}$, where $s_k$ and $c_k$ are the fitted coefficients of  the $\sin$- and $\cos$-terms, respectively, and 
the phase is defined as $\arctan(s_k/c_k)$. This procedure allows us to improve on the knowledge of $P_e$ using the entire
collection of data, because a slight offset in this parameter results in a clearly detectable linear trend of the measured phase.
The least-squares fit is generated for 29 consecutive bins of flux measurements, each spanning 6.79 d. A zero-point epoch
$t_0=1339.35$ (barycentric Julian day minus 2454833.0) 
is chosen to produce zero phases (on average) for the dominant harmonic $\sin(2\pi(t-t_0)/P_e+\phi)$, i.e., at
this time the main Fourier term crosses zero while increasing. 

\begin{figure}[htbp]
  \centering
  
  \includegraphics[width=2.15in]{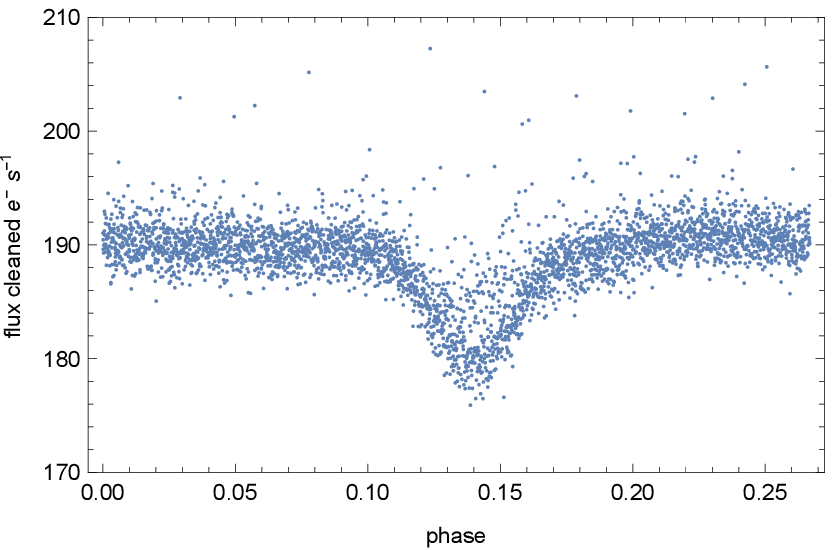}
  \includegraphics[width=2.15in]{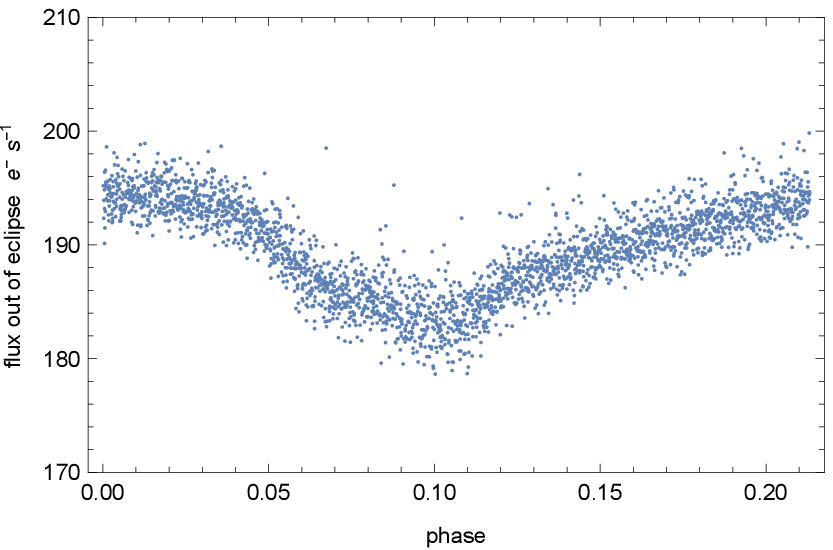}
  \includegraphics[width=2.15in]{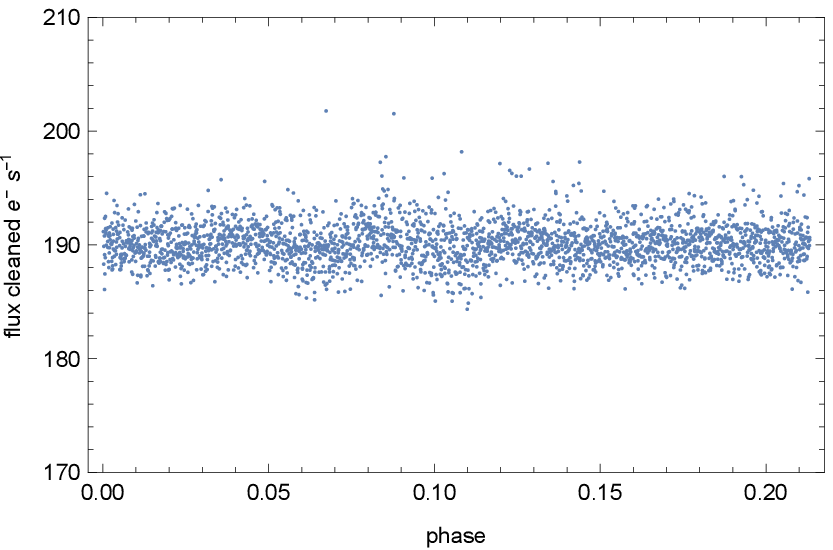}
  \caption{Folded light curves of \st\ in quarter 16. Left: Light curve with 6 sinusoidal harmonics of
  the $P_e$ variation filtered out, folded with the orbital period $P_o=0.2668707$ d. Center: Out-of-eclipse
  segments of the light curve folded with the period $P_e=0.213236$ d. Right: Out-of-eclipse segments
  of the light curve with 6 sinusoidal harmonics of
  the $P_e$ variation filtered out, folded with the period $P_e=0.213236$ d.}
  \label{clea.fig}
\end{figure}

The resulting phase determinations for the first harmonic ($k=1$) show a remarkably flat curve, i.e., no time dependence, and
a standard deviation of $2.5\degr$ around the mean of $-0.1\degr$. 
This precludes a more distant and sufficiently massive orbiting companion with periods
up to $\sim1$ year, which would have caused a measurable light travel time (LTTE) effect. The improved period based on
the
three quarters of data is $P_e=0.213236(2)$ d. The phases of the second harmonic show no time dependence either, but are of more
modest precision, grouping around $80.8\degr$ with a std of $10.7\degr$. This value means that the term is close to its maximum
value at the chosen reference epoch. The fitted Fourier harmonics and the mean flux can be subtracted from the entire light curve
(including the eclipse segments), and then the data can be time-folded with the $P_o$ period. The result for Q16 is shown in
Fig.~\ref{clea.fig}, left. A remarkable improvement in the scatter of data points is obvious, and the shape of the eclipse,
as well as the out-of-eclipse behavior, are more visible now. The eclipse is distinctly triangular, without a flat bottom,
which hints at a companion of comparable radius and a tangent, partial eclipse. Fig.~\ref{clea.fig}, center, shows the out-of-eclipse
part of the light curve for Q16 folded with $P_e$ and the same reference epoch. We note that there are times, separated by
$5\,P_e$, when the eclipse minimum nearly coincides with the main sinusoid minimum. At these instances, the eclipsing 
companion is the closest to the line of sight, and the periodic part of the flux is the dimmest. The cleaned out-of-eclipse
part of the light curve folded with $P_e$ (Fig.~\ref{clea.fig}, right) is flat with a std scatter of 1.8 e$^{-}$ s$^{-1}$.

\subsection{Tidal deformation caused by a tertiary companion}
Many eclipsing stars with periods shorter than 1 day show signs of ellipsoidal variation caused by the tidal deformation
of the orbiting companions. The perturbing tidal potential may have several significant modes (depending mostly on the
relative orbital separation and the eccentricity), but the greatest mode for nearly circular orbits is semidiurnal, 
whose frequency is exactly twice the orbital frequency. For \st, the presence of an orbiting companion is witnessed by the
periodic eclipse, but it can not be responsible for the sinusoidal variation because of the unmatched period. The only
remaining possibility is a third component of the system orbiting the primary with a period of $2\,P_e$ and generating
the observed ellipsoidal light curve. 

This explanation meets a few crucial difficulties. The orbits of the hypothetical tertiary and the eclipsing companions should
be in the $8/5$ period commensurability, i.e., in a mean motion resonance (MMR). Although $8/5$ MMRs have been
detected for exoplanet systems \citep{bar, gil}, they seem to arise when more than two companions are present
in a tightly packed configuration, and their masses are much smaller than the mass of the star. In this case,
a significant stellar mass is required to produce ellipsoidal variations of the observed amplitude, which makes
long-term dynamical stability of the system questionable.

The relative flux variation in each of the tidally interacting stars can be written as \citep{mor}:
\eb
F/F_0=\bar F + q\sum_{n,k}\alpha_{nk} \left(\frac{R}{a}\right)^n f_{nk}(i)\, \cos k\phi
\label{ell.eq}
\ee
where $F_0$ is the flux from a non-perturbed star, $\bar F$ is the mean disk-integrated, relative flux from a perturbed star,
$\alpha_{nk}$ are coefficients, which depend on the limb-darkening and gravity-darkening parameters, $f_{nk}$ are functions
of $i$, which is the inclination of the orbital axis to the line of sight, and $\phi$ is the perturber's longitude with respect to
the line of sight (zero at the eclipse minimum). For detached systems, $R/a$ is small, and only a limited number of terms
may be of interest. The coefficients and inclination functions of all significant terms are listed in Table~1. $R$ is
the radius of the perturbed star, and $q$ is the ratio of the perturber's mass and the mass of the perturbed star. The lowest
power of $R/a$ in the 32 semidiurnal term marks it as by far the largest. For \st, the observed amplitude of the main harmonic is
approximately 30 ppt, and the estimated $R/a$ is greater than 3 (but can be as large as 6). The mass of the tertiary
companion then would be comparable to, or several times greater than the mass of the eclipsed M dwarf, which seems
improbable.

Perhaps, a more specific test comes from the consideration of estimated amplitudes and phases of the periodic signal.
Table~1 suggests that the semidiurnal harmonic should be much greater that the other terms, if detectable at
all. Interpreting the strongest periodicity as the 32 term (proportional to $\cos 2\phi$), we also find a clearly
present $P_e/2$ harmonic, which would be the 54 term, but no sign of the 41 or 43 terms. The former can vanish
if the inclination is close to $63.43\degr$ due to its inclination function, but the latter should always be
larger in amplitude then the 54 term. However, we note that Eq.~\ref{ell.eq} and Table~1 are derived for a circular
orbit. Deformations caused by an eccentric perturber can have a different distribution of power between the harmonics
with the peak shifted toward higher orders of the perturbing potential. But this implies an exotic triple system
with a stellar-mass outer companion in a rather eccentric orbit, in which the eclipses are caused by a smaller
mass inner companion, such as a bloated giant planet.

Finally, the signs of $\alpha_{32}$ and $\alpha_{54}$ are opposite (Table~1). This means that the maxima of these
two harmonics should coincide when $\phi=\pi/2$, but the minima cannot coincide. On the contrary, the observed
phases of these two terms suggest that their minima almost overlap at times separated by $P_e$, which is
characteristic of eclipses but not of ellipsoidal variation.

\subsection{Yin-yang stars}
M dwarfs display a wide range of rotation periods, which are correlated with their age and activity level. The majority
of old, inactive stars rotate slowly, with insignificant consequences for the light curve. The system of \st\ is
moderately active, as witnessed by the frequent flares, which is expected of a close binary. It can not be precluded
that one of the components rotates fast and modulates a light curve with the observed period $P_e$. Stars like the Sun
have surface brightness irregularities, such as spots and plages, that are small-scale and short-lived. The observed
photometric effects are stochastic and low-amplitude \citep{mak09}, making their detection a difficult task.
On the other hand, active dwarfs in binary systems may have large surface structures occupying a significant fraction
of the visible disk. Presumably, they are also long-lived and appear in the line of sight for many consecutive
rotation periods, generating a strong modulation signal. 

The amplitude of a single spot modulation in relative flux is proportional to $(1-f_s)r_s^2$, where $f_s$ is the
contrast of the spot (close to unity) and $r_s$ is the central angle spanned by the spot's radius in radians. 
The deficit (or excess) of flux
in the spot, $1-f_s$ is a small positive or negative number, which is not expected to exceed several percent in
absolute value. A rough estimation reveals that in order to generate a 3\% modulation observed for \st, the radius
of the spot structure should approach $\pi/2$, i.e., the structure should fill almost half of the surface. In
this extreme scenario, the star is painted different colors on two sides. Like with the ellipsoidal variation
scenario, the first crucial problem is that the observed sinusoidal variation is too large.

The main mode of a surface structure modulation is close to the equatorial rotation period within the spread caused
by the differential rotation and the latitude. The nearly perfect $5/4$ commensurability between $P_o$ and $P_e$
implies that the star rotates by exactly 25\% faster than the occulter's orbital motion. This condition is called
spin-orbit resonance, and it is a natural consequence of the frequency-dependent secular tidal torque for
solid planets \citep[e.g.,][]{efr, mae}. Stars, however, are closer to the ``semiliquid" regime where the main
viscous response to tidal perturbation is greatly smoother in frequency, and the absence of a permanent shape
results in the pseudosynchronous stable equilibrium \citep{mur, mak15}. In that state, the perturbed body rotates
faster than the synchronous rate, and the offset in frequency is a smooth function of the orbital eccentricity. There
is no condition that forces the equilibrium rate to be in exact integer ratio to the mean motion.

On the remote chance that the star has an unexpectedly high viscosity and a finite rigidity (caused by magnetic turbulence,
for example), we can consider well understood planetary resonances. The tidal modes in this case split into a spectrum
defined by 4 integer indices $lmpq$:
\eb
\omega_{lmpq}=(l-2p+q)\,n-m\,\Omega,
\ee
where $n=2\pi/P_e$ is the mean motion, $\Omega$ is the spin rate, $l=2,3,\ldots$, $p=0,\ldots, l$, $m=0,\ldots, l$,
and $q$ is any integer between $-\infty$ and $+\infty$. A spin-orbit resonance occurs when $\omega_{lmpq}=0$.
$m$ is the number of sectorial spherical harmonics in the perturbation potential, which is limited to the degree $l$.
The lowest degree of a $\Omega/n=5/4$ spin-orbit resonance is then $l=4$. However, the degree-4 tidal perturbation should be
vanishingly small compared with the main $l=2$ (quadrupole) perturbation for well detached systems, because the 
potential is proportional to $(R/a)^{l+1}$ \citep[e.g,][]{kum,pfa} for a given tidal mode. We conclude that
photospheric structures and a tidal spin-orbit resonance can hardly explain the detected photometric signals.

\subsection{Heartbeat stars}
Non-radial pulsations can be perpetually excited by close companions in eccentric orbits and show up as quasi-sinusoidal
variations at frequencies in integer-number commensurabilities with the orbital frequency. KOI-54 is perhaps the
most impressive example of the class with sharp brightening peaks in the light curve separated by an orbital
period of 41.8 d and tidally excited non-radial pulsations with frequencies that are exact 90th and 91st integer multiples
of the orbital frequency \citep{wel}. The class of eccentric heartbeat binaries is characterized by orbital
periods of 4--20 d and often pulsations at higher frequencies \citep{tho}. As with KOI-54, the pulsation frequencies
are large multiple integers of the tidal frequency. They show up in the time-folded light curves as wiggles overlaying
the eclipsing and, sometimes, ellipsoidal variations. HD 183648 is an interesting example of a low-eccentricity
heartbeat binary with an out-of-eclipse semidiurnal periodicity, which seems to have the opposite phase to the
expected ellipsoidal modulation \citep{bor}. One may wonder if the out-of-eclipse modulation for HD 183648 is also a
heartbeat pulsation in a 2:1 resonance with the orbital period, and its observed phase is simply caused by the
periastron direction being at roughly the right angle from the line of sight. Other cases of detected period
commensurability include the $6/1$ ratio in the HAT-P-11 planetary system \citep{bek}, which is interpreted as
a spin-orbit resonance and a persistent surface spot, and a 5/3 ratio for the KOI-13 system \citep{sza}, also
interpreted as a spin-orbit resonance. We discussed in the previous paragraphs why this interpretation is
unlikely.

 \begin{table*}
 \centering
 \caption{Theoretical coefficients and inclination functions of ellipsoidal flux variations for the $V$ and $R$ photometric bands and a circularized orbit.}
 \label{coe.tab}
 \begin{tabular}{@{}lrrr@{}}
 \hline
            &          &       &      \\
      &  $f(i)$ & $V$    & $R$ \\
 \hline
 $\alpha_{32}$ & $\sin^2 i$ & $-1.21$ & $-1.14$  \\
 $\alpha_{41}$ & $4\sin i-5\sin^3 i$ & $+0.12$ & $+0.09$  \\
 $\alpha_{43}$ & $\sin^3 i$ & $-0.21$ & $-0.15$  \\
 $\alpha_{52}$ & $6\sin^2 i-7\sin^4 i$ & $-0.21$ & $-0.22$  \\
 $\alpha_{54}$ & $\sin^4 i$ & $+0.37$ & $+0.39$  \\
 \hline
 \label{table}
 \end{tabular}
 \end{table*}

\citet{rod12} suggested possible mechanisms of pulsation for M dwarfs and conditions of their detection. One
of the mechanisms is expected to produce pulsations in the period range of interest, i.e., a few to several
hours depending on the mass, but it is caused by the strong temperature dependence of deuterium burning,
and therefore, applies only to very young stars. Fully convective old stars of low mass can pulsate due to
He$^3$ burning, but the period is expected to be shorter than 41 min. Attempts to detect these pulsations
have been unsuccessful \citep{rod, rod16}.

\section{Summary}

\citet{gai} estimated for \st\ a radius of $0.18 R_{\sun}$ and a mass of $0.18 M_{\sun}$ with a large uncertainty.
In the light of Gaia DR2 data, the star is much more distant and more luminous, with a mass of $\sim 0.4 M_{\sun}$.
\citet{arm} estimated close effective temperatures of the two companions in the range 3500--3600 K, but a radius
ratio of $0.5\pm0.4$. The paucity of accurate data leaves room to speculation about the nature of this binary.
The eccentricity of the system can be significant, based on the absence of secondary eclipses, but perhaps not
very high, because no periastron brightening is present either. We considered three possible scenarios for the
discovered photometric commensurable periodicities and found two of them to be rather outlandish. An ellipsoidal
tidal deformation does not pass because it requires a stellar-mass tertiary component in a 8/5 MMR, and the amplitudes
and phases of the harmonics do not match the expectation. A persistent photospheric spot would have to be very large
or exceptionally dark to explain the observed amplitude of the main mode, and the rate of rotation would need to be
in a nearly perfect 4/5 spin-orbit resonance, for which we do not see a theoretical justification.

This leaves us with the possibility of a non-radial pulsation excited by tidal interaction. The model by \citet{kum}
of the class now known as heartbeat stars predicts a large morphological variety of signals driven by the
multitude of geometrical and orbital configurations. This model has often (but not always) been successful in describing
the periastron impulse-like variations in Kepler light curves \citep{tho}. In \st, we may be seeing a new type
of heartbeat stars without prominent periastron impulses but with pulsations that are tightly spaced from the excitation
frequency. It is possible that the damping time is short for the proposed mechanisms and they become prominent
only when excited by a variable tidal force in sufficiently close binaries. Such binaries are usually 
active and flaring, further concealing the intrinsic pulsation. Interestingly, we find an additional strong periodicity
at the high-frequency end of the available spectrum with a period of 0.022597 d $=$ 32.54 min and an amplitude of
3.2 e$^-$ s$^{-1}$, but it cannot be validated being too close to the long cadence observation frequency.
But the 5.1-hr signal we discovered does not currently have a theoretical background as an intrinsic non-radial pulsation.

\label{sum.sec}
\section*{Acknowledgments}
This paper includes data collected by the Kepler mission. Funding for the Kepler mission is provided by the NASA Science Mission directorate.
Some of the data presented in this paper were obtained from the Mikulski Archive for Space Telescopes (MAST). STScI is operated by the Association of Universities for Research in Astronomy, Inc., under NASA contract NAS5-26555. Support for MAST for non-HST data is provided by the NASA Office of Space Science via grant NNX09AF08G and by other grants and contracts. 
This research has made use of the VizieR catalogue access tool, CDS,
Strasbourg, France. The original description of the VizieR service was
published in A\&AS 143, 23.

\label{lastpage}

\end{document}